\documentclass[review]{elsarticle}
\graphicspath{ {./figures/} }
\usepackage{hyperref}
\usepackage{float}
\usepackage{verbatim} 
\usepackage{apalike}
\usepackage{enumitem}
\usepackage{xcolor}
\usepackage{soul}
\usepackage{booktabs}
\usepackage{adjustbox}
\usepackage{colortbl}
\definecolor{mygray}{rgb}{0.8,0.8,0.8}

\usepackage{subcaption}
\usepackage{algorithm}
\usepackage{algpseudocode}

\restylefloat{figure}
\restylefloat{table}

\journal{Expert Systems with Applications}

\biboptions{authoryear}

\begin{document}
\begin{frontmatter}

\begin{titlepage}
\begin{center}
\vspace*{1cm}

\textbf{ \large Use of recommendation models to provide support to
dyslexic students }

\vspace{1.5cm}

Gianluca Morciano$^{a}$ (gianluca.morciano@unitus.it), José Manuel Alcalde Llergo$^a$ (jose.alcalde@unitus.it), Andrea Zingoni$^a$ (andrea.zingoni@unitus.it), Enrique Yeguas Bolívar$^b$ (eyeguas@uco.es), Juri Taborri$^a$(juri.taborri@unitus.it), Giuseppe Calabrò$^a$(giuseppe.calabro@unitus.it)\\

\hspace{10pt}

\begin{flushleft}
\small  
$^a$ Department of Economics, Engineering, Society and Business Organization (DEIM), University of Tuscia,
01100 Viterbo, Italy \\
$^b$ Department of Computing and Numerical Analysis, Córdoba University, 14071 Córdoba, Spain

\vspace{1cm}
\textbf{Corresponding Author:} \\
Gianluca Morciano \\
Department of Economics, Engineering, Society and Business Organization (DEIM), University of Tuscia,
01100 Viterbo, Italy\\
Email: gianluca.morciano@unitus.it

\end{flushleft}        
\end{center}
\end{titlepage}

\title{Use of recommendation models to provide support to dyslexic students}

\author[label1]{Gianluca Morciano \corref{cor1}}
\ead{gianluca.morciano@unitus.it}

\author[label1]{José Manuel Alcalde Llergo}
\ead{jose.alcalde@unitus.it}

\author[label1]{Andrea Zingoni}
\ead{andrea.zingoni@unitus.it}

\author[label2]{Enrique Yeguas Bolívar}
\ead{eyeguas@uco.es}

\author[label1]{Juri Taborri}
\ead{juri.taborri@unitus.it}

\author[label1]{Giuseppe Calabrò}
\ead{giuseppe.calabro@unitus.it}

\cortext[cor1]{Corresponding author.}
\address[label1]{Department of Economics, Engineering, Society and Business Organization (DEIM), University of Tuscia,
01100 Viterbo, Italy }
\address[label2]{Department of Computing and Numerical Analysis, Córdoba University, 14071 Córdoba, Spain}

\begin{abstract}
Dyslexia is the most widespread specific learning disorder and significantly impair different cognitive domains. This, in turn, negatively affects dyslexic students during their learning path. Therefore, specific support must be given to these students. In addition, such a support must be highly personalized, since the problems generated by the disorder can be very different from one to another. In this work, we explored the possibility of using AI to suggest the most suitable supporting tools for dyslexic students, so as to provide a targeted help that can be of real utility. To do this, we relied on recommendation algorithms, which are a branch of machine learning, that aim to detect personal preferences and provide the most suitable suggestions. We hence implemented and trained three collaborative-filtering recommendation models, namely an item-based, a user-based and a weighted-hybrid model, and studied their performance on a large database of 1237 students' information, collected with a self-evaluating questionnaire regarding all the most used supporting strategies and digital tools. Each recommendation model was tested with three different similarity metrics, namely Pearson correlation, Euclidean distance and Cosine similarity. The obtained results showed that a recommendation system is highly effective in suggesting the optimal help tools/strategies for everyone, with an error less then 12\%. As a further evidence of the effectiveness of the implemented system, its precision was 0.85 and its recall was 0.83. The best performing filter was the hybrid one, when Pearson's correlation is used to measure the distance among users and/or items. In addition, in a final testing performed on 50 students, dyslexic students who used the recommendation algorithm increased their academic scores of almost 1 point in a 1 to 10 scale, showing higher learning performance compared to students who did not use it. This demonstrates that the proposed approach is successful and can be used as a new and effective methodology to support students with dyslexia. 
\end{abstract}


\end{frontmatter}

\section{Introduction}
\label{introduction}
Dyslexia is a neurodevelopmental disorder that can negatively impact several cognitive abilities that are crucial for learning. Therefore, a timely diagnosis and the use of valid support systems are of outmost importance in order to help the children to cope with their difficulties. The diagnosis of dyslexia can be done by using standard pen-and-paper tests \citep{schulte2010prevention} or by using machine learning (ML) algorithms. An example of the use of ML to diagnose dyslexia can be found in \citep{palacios2016extension}.
After the diagnosis, however, the use of effective support items is crucial to help the children to overcome some of their difficulties. 
ML and virtual reality (VR) represent ways in which support methods can be created and used by dyslexic children. For instance, in {\citep{maresca2022use}} VR was used to enhance cognitive abilities in dyslexic children while in {\citep{metroxraine2022}} was used to determine the difficulties of each tested student. ML and, in particular, recommendation systems are used to customize the learning experience of students with or without cognitive difficulties {\citep{urdaneta2021recommendation, hoic2015recommender,rodriguez2015student}}. Our work compares three recommendation systems using three different similarity metrics. The goal is to select the best model to assist university students with dyslexia in choosing the most effective learning tools and studying strategies for their needs. To provide a comprehensive understanding, the following sections will delve into the relevant literature (Section \ref{Related works}) and the methodologies employed in this study (Section \ref{methods}). Subsequently, Section \ref{results} will present the results, leading to a succinct conclusion in Section \ref{conclusion} summarizing the key insights gained from our investigation.

\section{Related works}
\label{Related works}
Specific learning disorders (SLDs) are a set of neurodevelopmental disorders that affect around 10\% of the worldwide population \citep{narimani2016compare} and impair, to different degrees, specific cognitive skills such as reading, writing and performing mathematical operations correctly \citep{shaywitz1998dyslexia}. Depending on the severity and the type of the impaired skill, three categories of SLDs can be distinguished \citep{toffalini2017strengths}: (i) dyslexia, if the subject has strong difficulties in reading; (ii) dysgraphia, when impaired writing skills are predominant and (iii) dyscalculia, if the mathematical skills are the most affected. In addition, there is a high rate of comorbidity between these SLDs, thus often two or more disorders are present together. Among them, dyslexia has the highest prevalence \citep{wagner2020prevalence}. Dyslexia affects not only reading but also attention \citep{bosse2007developmental}, working memory \citep{smith2007working} and phonological processing \citep{galaburda2006genes}. Also, abilities such as note-taking, text composing and organizing the study activity are negatively influenced \citep{mortimore2006dyslexia}. These deficits can significantly impact the learning experience of dyslexic students. However, numerous studies show that if dyslexia is detected early during childhood, it is possible to reduce the severity of the symptoms by employing supporting methodologies that help in addressing the specific difficulties of the child \citep{battistutta2018impact}. Therefore, diagnosing dyslexia as early as possible is of paramount importance. 

Among the methods used to diagnose dyslexia, the most common are the cognitive tests \citep{schulte2010prevention}. These, are used to examine general cognitive abilities, using the Wechsler Adult Intelligence Scale (WAIS) for adults or the Wechsler Intelligence Scale for Children (WISC) \citep{wechsler2014wisc} for children, or to assess specific cognitive abilities such as verbal and phonological processing \citep{pothos2004investigating}, generally compromised in dyslexia.

If the diagnosis of dyslexia does not arrive early in life, during childhood or adolescence, or if it fails, the difficulties produced by dyslexia tend to be more accentuated and acting to mitigate them becomes even more difficult \citep{battistutta2018impact}. In addition, these issues can influence self-esteem and future aspirations \citep{ghisi2016socioemotional} which increase the risk of developing generalized anxiety \citep{mammarella2016anxiety} or depression \citep{mugnaini2009internalizing}. In light of this, finding support tools for dyslexic students is of utmost importance. This support can come in the form of standard tools or digital tools or a combination of those. For instance in \citep{maccullagh2017university}, the authors interviewed dyslexic students about the support tools perceived as most useful. The students reported that one of the most effective standard support tool was attending face-to-face lectures and recording them. This was best coupled with the use of visual slides. 

An example of digital tool that is currently studied to help dyslexic students is the VR \citep{ metroxraine2023}. For instance, in \citep{maresca2022use} dyslexic students were divided and trained through classical neuropsychological treatments or using VR. The VR treatment was composed of different exercises aimed at improving cognitive functions such as memory, attention, language, spatial–temporal orientation and executive functions. The results show that the students that underwent the VR treatment improved their abilities across all the cognitive domains analyzed. This study shows the potential of using VR in supporting dyslexic students. 

However, even if providing standard and digital support tools helps dyslexic students, this may not be enough. As highlighted before, dyslexia has a high comorbidity with other SLDs. In addition, dyslexia has a comorbidity with other neurodevelopmental disorders such as attention deficit hyperactivity disorder (ADHD) \citep{germano2010comorbidity}, executive functioning disorder \citep{gooch2011time} or obsessive compulsive disorder \citep{pauc2005comorbidity}. Therefore, the presence of two or more disorders translates in two or more cognitive deficits co-existing and this leads to increased issues during the learning path of each student. Furthermore, greater deficits in learning can result in higher risk of developing psychological issues during childhood and adolescence. In light of this, it appears evident that different dyslexic students can have different difficulties and thus, the support strategies and tools designed to help them during their learning path must be personalized. With the recent advancements in machine learning (ML) algorithms, today it is possible to develop supporting methodologies that can ameliorate the difficulties encountered by dyslexic subjects during their life, especially during their education path \citep{ciolacu2017education},  \citep{baglama2018using}, \citep{zingoni2024machine}. For instance, in \citep{hamid2015computer}, researchers use a hidden Markov model to predict challenges in learning the Malay language among primary school students with dyslexia. The model analyzes errors in phonology, spelling, reading, and writing exercises, emphasizing the necessity for personalized support interventions. Similarly, \citep{mpia2013intelligent} adopts a comparable approach but incorporates students' behavior into the model training. The outcome is a computer-assisted learning system that, based on predictions, leads to a substantial 60\% improvement in dyslexic students' skills compared to traditional support tools. Meanwhile, the primary goals of \citep{rajapakse2018alexza,thelijjagoda2019hope} are focused on creating an assistive learning platform tailored for both primary and secondary school students, by using object character recognition and a neural network (NN) in the former or a support vector machine (SVM) in the latter.

Among the ML algorithms, recommendation systems (RS) are a natural candidate to provide a personalized learning experience to students \citep{masruroh2021adaptive} as they are specifically designed to detect user preferences and recommend items accordingly \citep{khalid2022literature}. RSs can be generally categorized in collaborative filtering (CF), in which items are recommended based on the preference of similar users \citep{herlocker2004evaluating}, content-based (CB), in which items are suggested to the user based on the description of the item \citep{marana2023rating}, knowledge-based (KB), which focuses on the knowledge of the users' need for a particular item \citep{carrer2012social} and hybrid system which is a combination of the algorithms described above \citep{burke2002hybrid}. In the field of education, several RSs were proposed to enhance the learning experience of the students \citep{nabizadeh2020learning} . For instance, in \citep{hoic2015recommender} the authors combined a set of RSs such as knowledge-based, CB and CF along with Web 2.0 tools to provide teachers with suggestions on how to design a course and students with indications on what activities to choose such as attending seminars, participating in e-courses, summaries of subject matter and solving online tests. Thanks to this method, students achieved better final course results. In \citep{rodriguez2015student} the authors created an hybrid model in which a CF, a CB and a KB algorithm were combined in order to recommend learning objects to students that have similar learning styles. This hybrid recommendation approach increased the relevance of the recommended educational material and improved the students' learning process. 

The studies reported above and in Table \ref{tab:table_references} show a prominent role of RSs in education. 
\begin{table}[b!]
\centering
\begin{adjustbox}{width=\textwidth}
\caption{ Brief overview of ML studies aimed at helping students with and without learning difficulties during their academic education. The acronyms in the table refer to different ML models. In particular, NN: Neural network, OCR: object-character recognition, SVM: Suppor Vector Machine, k-NN: k-Nearest Neighbors, LR: Logistic Regression, RF: Random forest}
\label{tab:table_references}
\begin{tabular}{lll}
\hline
\multicolumn{1}{|l}{Reference}                  & Type of ML                                                   & \multicolumn{1}{l|}{Description}                                                                                                                                                                            \\ \hline
\citep{hamid2015computer} & Hidden Markov & \begin{tabular}[c]{@{}l@{}}ML tracks children's mistakes \\
in phonology,speling, reading and writing\end{tabular} \\ \hline
\citep{hoic2015recommender}    & CB, CF, KB                                                   & \begin{tabular}[c]{@{}l@{}}RS systems were combined with\\ Web applications in order to help teachers \\ in design courses and students to choose \\ the activities to integrate in their learning path\end{tabular} \\ \hline
\citep{rodriguez2015student}   & CB, CF, KB                                                   & \begin{tabular}[c]{@{}l@{}}RS systems used to recommend learning \\ objects to student with similar learning styles\end{tabular}                                                                             \\ \hline
\citep{sergis2015enhancing}      & CF                                                           & \begin{tabular}[c]{@{}l@{}}RS is used to suggest, to teachers, new resources\\ to enhance their teaching practices\end{tabular}                                                                              \\ \hline
\citep{bourkoukou2016learning} & CF                                                           & \begin{tabular}[c]{@{}l@{}}RS used to select the best learning path based\\ on the learning style of each user\end{tabular}                                                                                  \\ \hline
\citep{meryem2016toward}       & Hybrid RS                                                    & \begin{tabular}[c]{@{}l@{}}RS aimed at helping the students in choosing their\\ career path\end{tabular}                                                                                                     \\ \hline
\citep{rajapakse2018alexza} & OCR and NN & \begin{tabular}[c]{@{}l@{}}OCR and NN used to support students while \\ reading difficult words\end{tabular} \\ \hline   
\citep{thelijjagoda2019hope} & SVM & \begin{tabular}[c]{@{}l@{}}SVM used to change the difficulty of video-games \\ and exercises.The students report improvements in \\ their skills and are more engaged in learning \end{tabular}                                                                                    \\ \hline
\citep{pan2020multiview}       & NN                                                           & \begin{tabular}[c]{@{}l@{}}Used NN to recommend courses to students and\\ identify those with learning risks\end{tabular}                                                                                    \\ \hline
\citep{suganya2020subjective}  & SVM                                                          & \begin{tabular}[c]{@{}l@{}}ML used to predict students performance and\\ to indicate directions for improvement\end{tabular}                                                                                 \\ \hline
\citep{zingoni2024machine}     & \begin{tabular}[c]{@{}l@{}}SVM, k-NN, \\ LR, RF\end{tabular} & \begin{tabular}[c]{@{}l@{}}ML models used to suggest the best digital\\ tools and learning strategies to dyslexic students\end{tabular}                                                                      \\ \hline
\end{tabular}
\end{adjustbox}
\end{table}
Unfortunately, the majority of RSs focus on students without SLDs such as dyslexia. Moreover, little to no effort is done in creating study support tools to dyslexic students at university except for \citep{zingoni2021investigating, zingoni2024machine} that mixes AI and virtual reality to create a platform to support dyslexic students.

Our work was developed with the goal of exploiting ML to customize the learning experience of students by recommending specific support tools. In particular, three RSs were compared to determine which algorithm recommend the best learning strategies and study tools according to the needs of each student. Lastly, the best performing algorithm has been used to recommend learning tools and study strategies to dyslexic and non-dyslexic students and their learning rate was assessed by experienced professors. 
This work is framed within the VRAIlexia project, which aims at provide specific support to university students with dyslexia, by relying on new technologies, like VR and AI \citep{vrailexia}.

\section{Methodology} 
\label{methods}
In this work, the goal was to compare the efficacy of three RSs, a user-based, an item-based and a hybrid model, in suggesting the best study tools and learning strategies to university students with dyslexia. Along with these RSs, three similarity metrics were tested as well namely, the Euclidean distance, the Cosine distance and the Pearsons' correlation. After the collection of the dataset, the hybrid model was implemented by combining the user-based and the item-based with different weights, as shown in \citep{badaro2013hybrid}. In order to select the best set of weights, the mean absolute error was computed, and subsequently, the algorithm's performance using these weights was further validated using other commonly used metrics, such as precision and recall at k. Finally, the best performing model was used to help students with and without dyslexia during their learning path.
\label{Material and Methods}
\subsection{Dataset collection}

\begin{figure}[ht]
  \centering
  \includegraphics[width=0.6\linewidth]{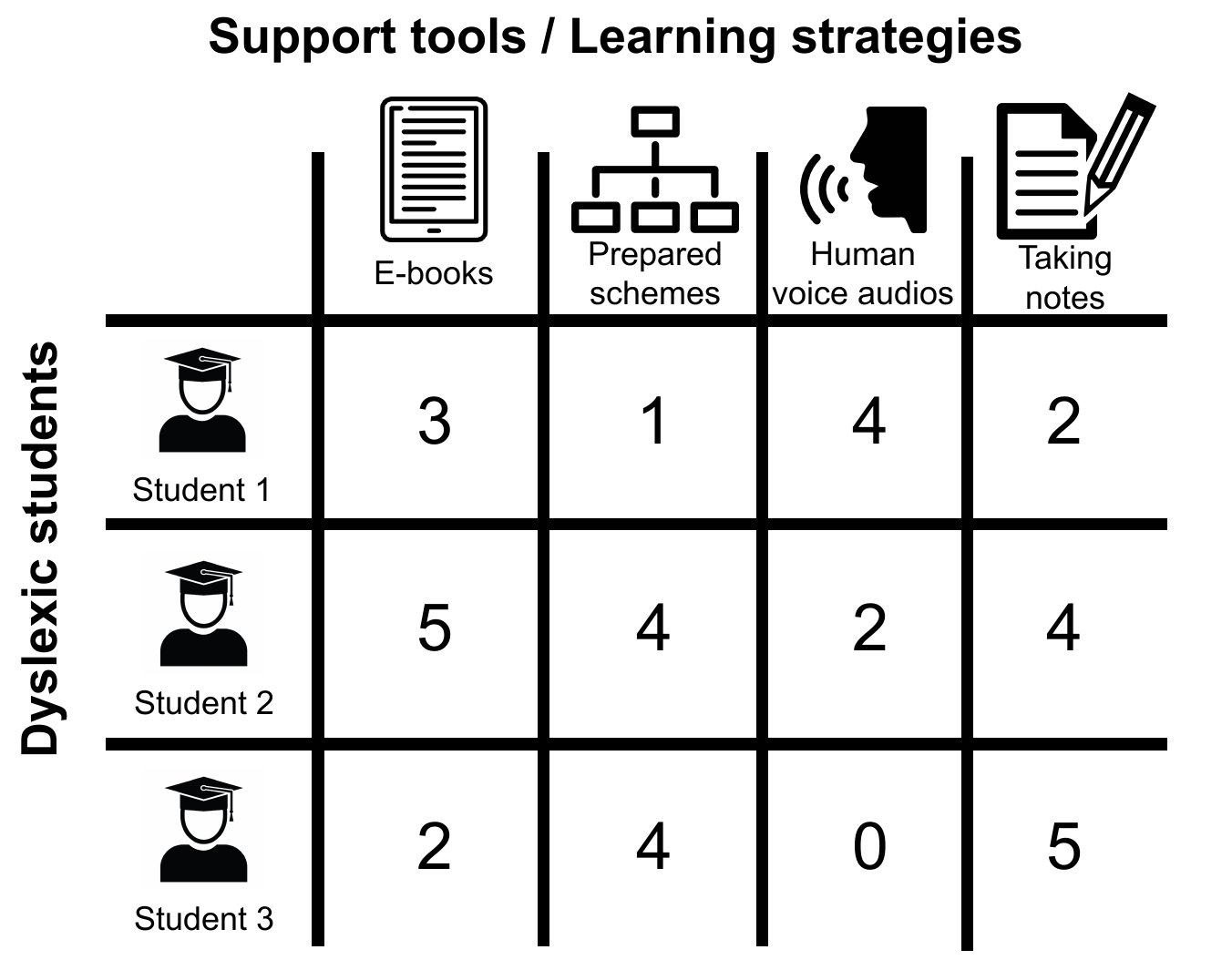}
  \caption{Rating matrix representation format.}
  \label{fig:matrix}
\end{figure}

The dataset used in this work was collected from a questionnaire aimed at assessing the issues that each dyslexic student has experienced and the learning strategies and study tools they have used and found useful to ameliorate their difficulties. Questions were also asked about the main problems caused by dyslexia, on the basis of which the most useful tools were able to be recommended in \citep{metroxraine2022}. In addition, demographic information has been collected in the dataset, but were used for other purposes such as shown in \citep{benedetti}. The different learning methodologies presented in the questions were selected with the help of experts in SLDs. Here, participants can describe how much the supporting tools and strategies presented were considered helpful by choosing different options: “not at all”, “very little”, “little”, “medium”, “much” and “very much”. However, to include the possibility that users never tried or did not know one or more items, two other answers (“never tried” and “I don’t know”) were added. These two responses were used to identify tools and strategies that were not sufficiently known to users, enabling the promotion of their application to obtain more accurate ratings in future analyses. In order to quantify a qualitative answer, and use it profitably in RSs, the different options have been converted into a range of numerical values between 0 and 5, were 0 corresponded to ``not at all''  and 5 to ``very much''. Regarding responses labeled as “never tried” and “I don’t know” they will be treated and preprocessed as missing values. The algorithm will process these data in the form of a ratings matrix, where the rows will represent different students, and the columns will represent support tools and study methodologies that may be proposed by the system. This representation follows a format similar to the one illustrated in Figure \ref{fig:matrix}encompassing all the students and the dyslexic support methodologies under consideration.

The questionnaire was created by skilled psychologists by starting from an interview carried out on 20 dyslexic university students about the best tools and strategies they have adopted to overcome those issues. The questionnaire was then administered to other 30 dyslexic students, in order to check for its accessibility and clearness. The final version, comprising 17 support tools and 22 learning strategies, totaling 39 items, was distributed to 1,237 dyslexic students, in accordance with the following criteria: having a valid diagnosis of dyslexia, being native Italian speakers, being 18 years old or more, and attending university or having finished or abandoned it less than five years before the filling of the questionnaire. 
The distribution of the participants was nearly even, concerning gender (54\% female students against 46\% male students) and age (an almost uniform distribution was observed within the selected range). In addition, the above-described information we collected about the received support, the type of high school attended, the category of student (full-time student, worker student, commuter etc.) and the family context allowed us to verify that all these variables were quite uniformly distributed. This limit potential biases. On the contrary, a surely present bias is given by the fact that only Italian-speaking participants took part to the questionnaire but, as reported previously limiting the work only to one language is strictly necessary, since dyslexia-related issues vary a lot from a language to another~\citep{Miles2000DyslexiaMS} and, thus, it is opportune to analyze each language by itself. It is worth mentioning that it was not possible to gather information concerning variables like the socioeconomic situation of the students and their ethnicity, which may have an impact on the system. However, at least the second one does not represent a great issue for the considered Italian university context, since this is less ethnically heterogeneous than other European countries, with only a 3\% presence of non-Italian students. The acquired data have been treated according to the article 13-14 of the GDPR 2016/679 \citep{parliament119council} of the European Union. In particular, they have been processed completely anonymously and used only for research purposes. Table \ref{tab:Tools_strat} presents the questions related to the issues, the tools and the learning strategies gathered by the questionnaire. However, it is important to note that the difficulties will not be taken into account when generating recommendations by the system.

\begin{table*}
\centering
\begin{adjustbox}{width=\textwidth}
\caption{Difficulties (P), support tools (T) and learning strategies (S) considered.}
\label{tab:Tools_strat}
\begin{tabular}{|c|c|c|c|}
\hline
\textbf{ID} & \textbf{Difficulty/Tool/Strategy}                                        & \textbf{ID} & \textbf{Difficulty/Tool/Strategy}                                       \\ \hline
P1          & Reading                                                    & T15         & Audio recording of lessons                                  \\
P2          & Writing                                                    & T16         & Video lessons                                               \\
P3          & Understanding difficult words                              & T17         & Supplementing study material with internet research         \\
P4          & Understanding the lessons                                  & S1          & A person reading for him/her                                \\
P5          & Concentration                                              & S2          & A map made by himself/herself                               \\
P6          & Paying attention during presential lessons                 & S3          & A scheme made by himself/herself                            \\
P7          & Paying attention during online lessons                     & S4          & A summary made by himself/herself                           \\
P8          & Memorising recently studied concepts                       & S5          & Repeat the studied material                                 \\
P9          & Remembering concepts studied during the exam               & S6          & Marking keywords                                            \\
P10         & Study time management                                      & S7          & Underlining with different colours                          \\
P11         & Taking notes                                               & S8          & Having a study group                                        \\
P12         & Limited time available to prepare a task/question/exam     & S9          & Having a tutor                                              \\
T1          & Human voice audio book                                     & S10         & Dyslexic student group to exchange resources                \\
T2          & Robotic voice audio book                                   & S11         & Presential lessons                                          \\
T3          & Different colour words                                     & S12         & Online lessons available                                    \\
T4          & Using the EasyReading font                                 & S13         & Taking breaks during lessons                                \\
T5          & Using a smart pen or tablet to take notes and record voice & S14         & Lesson slides available                                     \\
T6          & Clearer layout of the study material                       & S15         & Recording the lesson                                        \\
T7          & Having the key words of the text highlighted               & S16         & Taking notes                                                \\
T8          & Prepared concept maps                                      & S17         & Having the lesson plan in advance                           \\
T9          & Prepared schemes                                           & S18         & Dividing an examination/task/question into several parts    \\
T10         & Prepared summaries                                         & S19         & Only written tests                                          \\
T11         & E-Books                                                    & S20         & Only oral tests                                             \\
T12         & Digital tutor                                              & S21         & Conducting the exams in the presence of the professor alone \\
T13         & Images to help understand the meaning of difficult words   & S22         & Having an online database with notes made by other students \\
T14         & Images that help to memorise a concept                     &             &                                                             \\ \hline
\end{tabular}
\end{adjustbox}
\end{table*}

\subsection{Algorithms and metrics}

In recent years, RSs have been accruing increased significance, extending their influence into different fields such as education and learning. Among the various techniques employed in this domain, collaborative RSs and content-based approaches stand out as two distinctive methodologies, each with its unique strengths and limitations. Content-based systems require a big amount of information about items’ features, instead of focusing on the relationships obtained through feedback from users through ratings. For this reason, the collaborative systems approach has been considered more appropriate for our data set. Still, the content-based approach will be considered in future work to compare performance against collaborative methods, and ultimately provide the best possible advice to dyslexic students.

\begin{figure*}
    \centering
    \begin{subfigure}[b]{0.49\textwidth}
         \centering
         \includegraphics[width=\textwidth]{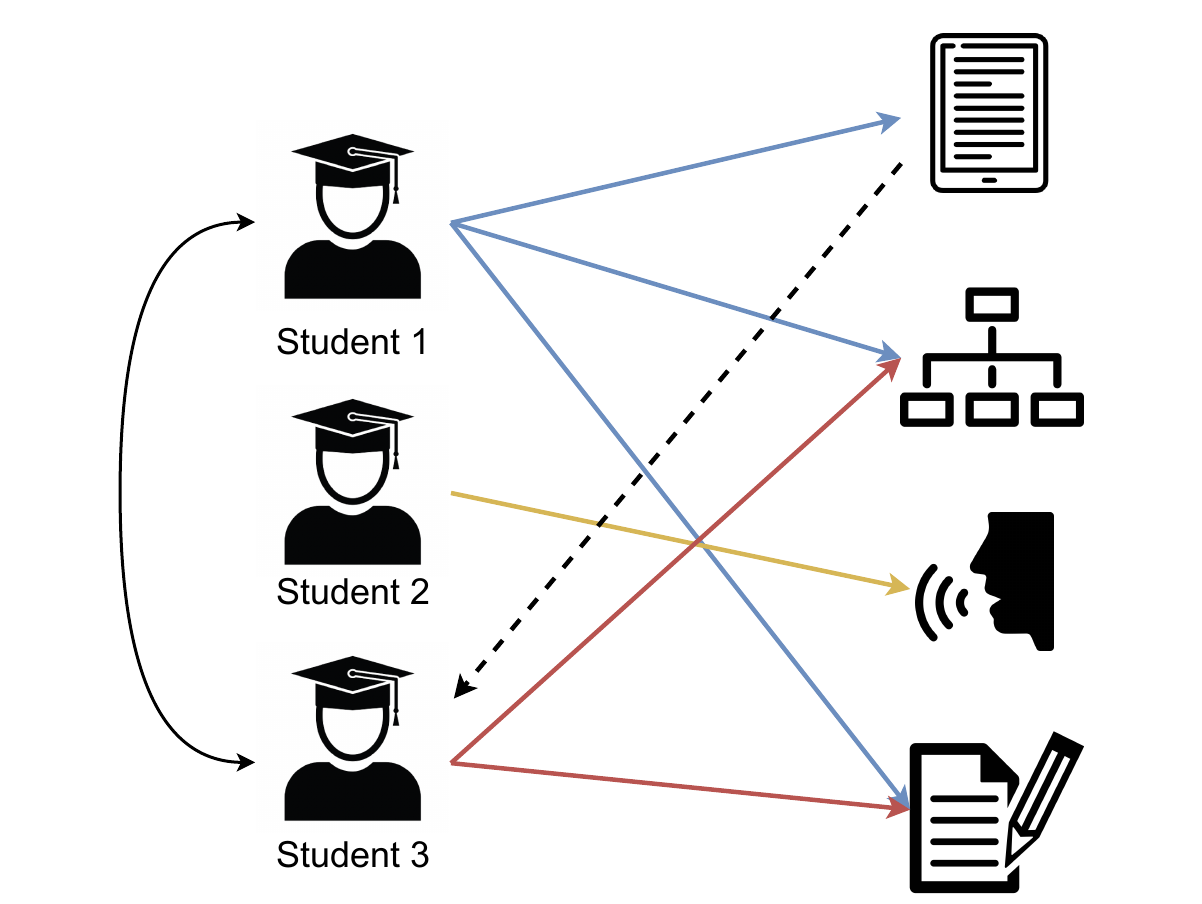}
         \caption{User-based collaborative filtering.}
         \label{subfig:userb}
    \end{subfigure}
    \begin{subfigure}[b]{0.49\textwidth}
         \centering
         \includegraphics[width=\textwidth]{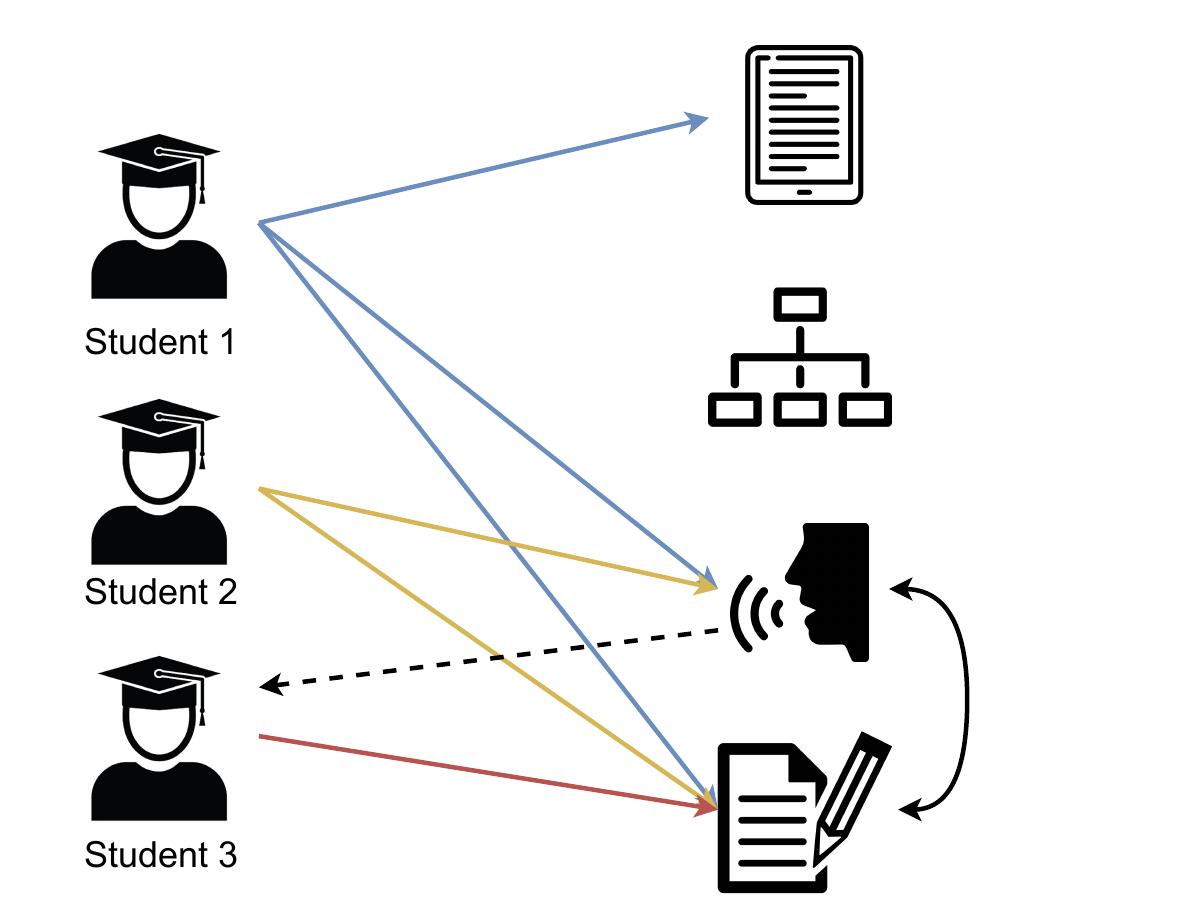}
         \caption{Item-based collaborative filtering.}
         \label{subfig:itemb}
    \end{subfigure}
    \caption{User-based approach (a) discovers that Student 1 and Student 3 are similar because they have chosen some similar learning methodologies, so it recommends (dashed arrow) to the Student 3 the item that was useful to Student 1 and that the latter is not using. Item-based approach (b) finds that the two last items have been rating similarly by different users (Student 1 and 2), so since Student 3 also likes one of these objects, it recommends the other one to the Student 3.}
    \label{fig:collab}
\end{figure*}

Data were analyzed using three types of CF RSs implemented as \citep{badaro2013hybrid}. The three types CF systems were: a user-based \citep{zhao2010user} in which the similarity between users guides the recommendation of the different methodologies, an item-based \citep{wang2006unifying} in which item similarity is analyzed to suggest the most relevant tools and strategies to the users, and a hybrid approach \citep{badaro2013hybrid} which combines the above mentioned approaches through a weight system. The motivation to use a hybrid model is given by the differences presented between the user-based and item-based approaches. User-based approach make recommendations based on the similarity between users, which means the algorithm will recommend items liked by users similar to the student. On the contrary, in the item-based approach recommendations are made based on similarities between items, computing this similarity according to the given ratings. The differences between the two approaches are visually depicted in Figure \ref{fig:collab}.

Hybrid approach was implemented by combining the user-based and item-based methods with different weights. This involved assigning specific relevance or weight to each of the predicted ratings using the equation below:

\begin{equation}
    \widehat{r} = \alpha\widehat{r}_u + (1-\alpha)\widehat{r}_i
\label{eq:hybrid}
\end{equation} 
 where $\alpha$ is the weight and $\widehat{r}_u$ is the predicted rating for the user-based approach, whereas $(1-\alpha)$ is the weight and $\widehat{r}_i$ is the predicted rating for the item-based algorithm.

 These weights were used to reduce the error deriving from predicted ratings and actual ratings as mentioned in \citep{badaro2013hybrid}. The value of $\alpha$ has to satisfy:

\begin{equation}
    \alpha \geq 0\; and \; \alpha \leq 1 \;
\end{equation}

To select the final weights to be assigned to each of the models, a preliminary experimentation has been carried out. After that, coinciding with the results obtained by \citep{sarwar2001item}, it was found that higher weights for the items-based method ($1-\alpha$) provided better results. Therefore, a scale of values for the weights has been selected to emphasize higher values for the item-based weight, starting with an user-based weight value ($\alpha$) twice as large as the item-based one, and concluding with item-based weight values seven times larger than $\alpha$. All the considered cases are illustrated in Table \ref{tab:weights}. 

\begin{table}[H]
    \centering
\begin{tabular}{|c|c|c|}
\hline
    Case & $\alpha$ & $(1-\alpha)$ \\
    \hline
    \#1 & $2/3$ & $1/3$ \\
    \hline
    \#2  & $1/2$ & $1/2$ \\
    \hline
    \#3  & $1/3$ & $2/3$ \\
    \hline
    \#4  & $1/4$ & $3/4$ \\
    \hline
    \#5 & $1/5$ & $4/5$ \\
    \hline
    \#6 & $1/6$ & $5/6$ \\
    \hline
    \#7 & $1/7$ & $6/7$ \\
    \hline
    \#8 & $1/8$ & $7/8$ \\
    \hline
\end{tabular}
    \caption{Configuration of weights for the hybrid approach.}
    \label{tab:weights}
\end{table}

It is worth noting that by assigning $\alpha=1$ or $\alpha=0$ the fully user-based approach and the fully item-based approach are obtained, respectively. Thus, by properly tuning $\alpha$ it is possible to include all the analyzed cases within the formula \ref{eq:hybrid}.

To compute similarities among users or items, three widely used metrics in CF systems were considered and compared: the Pearson correlation coefficient\citep{sheugh2015note}, the Euclidean distance and the Cosine distance \citep{vinh2018hyperbolic,hofmann2004latent}. These metrics aim to identify the n most similar users to the test user (in the case of user-based), or the n most similarly evaluated items according to the ratings given by the test user(in the case of item-based). These three metrics are defined as follow:

\begin{equation}
    Pearson\ Correlation=\frac{\sum_{i=1}^{N}(x_i-\bar{x})(y_i-\bar{y})}{\sqrt{\sum_{i=1}^{N}(x_i-\bar{x})^2}\sqrt{\sum_{i=1}^{N}(y_i-\bar{y})^2}}
\end{equation}

\begin{equation}
    Euclidean\ distance = \sqrt{\sum_{i=1}^{N}(x_i-y_i)^2}
\end{equation}

\begin{equation}
    Cosine\ distance = 1-\frac{\sum_{i=1}^{N}x_i y_i}{\sqrt{\sum_{i=1}^{N}x_i^2}\sqrt{\sum_{i=1}^{N}y_i^2}}
\end{equation}

where N is the number of users or items; $x_i$ and $y_i$ are the individual ratings; and $\bar{x}$ and $\bar{y}$ are the sample mean.

The Pearson coefficient is employed for direct calculation of similarities among all users or objects. Then, most similar neighbors are obtained by taking the n most correlated ones. In contrast, the Euclidean and Cosine distances are applied as distance metric within a K-Nearest Neighbors (k-NN) algorithm, and the n nearest neighbors are directly extracted from the obtained model. Another important aspect to take into account is the number of nearest neighbors considered in the similarity calculation. After conducting numerous tests with various potential values for the number of neighbors, we determined that optimal results are achieved with n values in the range of 3 to 11. Therefore, in the final experimentation, we will consider the number of neighbors to be 3, 5, 7, and 11. Algorithm~\ref{alg:sim} contains the pseudocode of the process for the similarity computation depending on the different considered metrics. The method returns the \textit{n\_neighbors} most similar users or items sorted according the computed similarity measure, stored in \textit{top\_similarities}.

\begin{algorithm}
\caption{Compute similarities}
\begin{algorithmic}[1]
\Function{ComputeSimilarities}{\textit{data\_similarities, similarity, n\_neighbors}}    
    \If{ \textit{similarity} $=$ ``pearson''}
        \State \textit{correlation\_matrix} $\gets$ correlation matrix computed using pearson coefficient
        \State \textit{test\_user\_corr} $\gets$ matrix row corresponding to the \textit{test\_user}
        \State \textit{similarities} $\gets$ sort similar according to \textit{test\_user\_corr}
        \State \textit{top\_similarities} $\gets$ first \textit{n\_neighbors} from similar\_users
    \Else  
        \State \textit{knn} $\gets$ generated k-Nearest Neighbors model using \textit{similarity} as distance metric and considering \textit{n\_neighbors} as the number of neighbors
        \State \textit{top\_similarities} $\gets$ nearest neighbors considered by \textit{knn}
    \EndIf
    
    \State \Return \textit{top\_similarities}
\EndFunction

\end{algorithmic}
\label{alg:sim}
\end{algorithm}

The objective of the final learning methodologies RS is to predict the ratings of tools and strategies for a particular user. The prediction is made by computing the average of the ratings given by the most similar users in the case of user-based approach or, in the case of item-based approach, the ratings of the most similarly scored items. In the case of the hybrid method, the prediction is obtained with the result of the sum of the user-based and object-based predictions multiplied by their corresponding weights.

The entire process described above is reflected in Algorithm~\ref{alg:user} and Algorithm~\ref{alg:item}, corresponding to the user-based and item-based recommendation systems, respectively. Both algorithms generate a list of key-value pairs, referred to as \textit{recommendations}, comprising the tools and strategies to be recommended along with their respective ratings assigned by the recommendation system. To achieve this, both algorithms distinguish the data they will utilize for similarity computation, \textit{data\_similarities}. In the case of user-based, only the training data, \textit{data\_train\_items}, and the test user's data, \textit{text\_user} for whom the recommendation is to be made, are needed. Conversely, in the case of item-based, it is necessary to consider each test item, from \textit{data\_test\_items}, separately to generate their respective ratings without taking into account the rest, treated as unknown to the user. Finally, \textit{recommendations} is completed with the average rating values obtained for the different test items from the \textit{n\_neighbors} most similar users or objects.

\begin{algorithm}
\caption{User-based collaborative filtering}
\begin{algorithmic}[1]

\Function{UserBasedRecSys}{\textit{test\_user, data\_train\_items, data\_test\_items, similarity, n\_neighbors}}
    \State \textit{recommendations} $\gets$ empty key-values map
    \State \textit{data\_similarities} $\gets$ concatenation of \textit{data\_train\_items} and \textit{test\_user} as row
    \State \textit{similarities} $\gets$ \Call{ComputeSimilarities}{\textit{data\_similarities, similarity, n\_neighbors}}
    \State \textit{recommendations} $\gets$ mean of ratings for each item taken from \textit{data\_test\_items}
    \State \Return 
\EndFunction

\end{algorithmic}
\label{alg:user}
\end{algorithm}

\begin{algorithm}
\caption{Item-based collaborative filtering}
\begin{algorithmic}[1]
\Function{ItemBasedRecSys}{\textit{test\_user, data\_train\_items, data\_test\_items, similarity, n\_neighbors}}
    \State \textit{recommendations} $\gets$ empty key-values map
    \For{\textit{test\_item} in data\_test\_items}
        \State \textit{data\_similarities} $\gets$ concatenation of \textit{data\_train\_items} and \textit{test\_user} as row
        \State \textit{data\_similarities} $\gets$ concatenation of \textit{data\_train\_items} and \textit{test\_item} as column
        \State \textit{similarities} $\gets$ \Call{ComputeSimilarities}{\textit{data\_similarities, similarity, n\_neighbors}}
        \State \textit{rating} $\gets$ mean of ratings given to the test item depending on train items
        \State add (\textit{test\_item}, \textit{rating}) to recommendations
    \EndFor
    \State \Return recommendations
\EndFunction

\end{algorithmic}
\label{alg:item}

\end{algorithm}

To assess the accuracy of the used methodologies, the mean absolute error (MAE) between the score predicted by RS and the respective score actually given by the user will be obtained. MAE is calculated by averaging all the $N$ absolute errors of the rating-pairs ($p_{i}$, $q_{i}$). Since lower MAE reflects a better accuracy the choice of the weights must be done in order to minimize it:

\begin{equation}
    \arg\min_{}(\alpha) (\frac{\sum_{i=1}^N \left |p_{i}(\alpha) - q_{i}(\alpha)  \right |}{N})
\end{equation}

where N is the total number of tools and strategies considered to be recommended, $p_{i}$ is the predicted value and $q_{i}$ the actual value.

Finally, one of the main challenges faced by RSs is the cold-start problem, which can be presented as a user cold-start or item cold-start \citep{sejwal2022hybrid}. The user cold-start problem stems from the challenge of addressing new users for whom the system lacks any pre-existing information \citep{userCS}. On the other hand, the item cold-start appears when a new item is being introduced to the dataset \citep{itemCS}. The former concern has been addressed through the solicitation of opinions from new users regarding the system's methodologies for profile generation, which are subsequently compared with those of other users. Regarding the item cold-start, it is not expected to pose a problem for our existing system. Indeed, new learning methodologies are not implemented very often due to this kind of strategies occasionally appears. The developed system will demonstrate its robustness by considering both test users and items when evaluating its performance.

\subsection{Data analysis}
From the 39 items used in the questionnaire, T4 support tool (Using the Easy Reading font) was discarded since more than 48\% of the participants did not know it. Therefore, the actual number of support tools used was 16, to which are added 22 learning strategies for a total of 38 items. The data were partitioned, with 75\% (947 users) utilized for training the model and the remaining 25\% (290 users) designated for the test set. Subsequently, a 10-fold cross-validation was applied to the training dataset to optimize the configuration parameters of the model before validating its performance on the test set. In addition, another separation of the data has been carried out to get some test learning methodologies and use them for the evaluation of the system. This split has been done taking randomly a 20\% of the learning methodologies along different epochs to check the robustness of the system in making predictions for different combinations of unknown items.
 
\subsection{Assessment} \label{section4}
Finally, to test the algorithm on a real case, the best performing model found was used to assess the efficacy
of recommending specific learning tools and strategies to students with or without dyslexia. 50 subjects, 53\%
male and 47\% female, participated to the testing campaign. Among them, 40\% are dyslexic, whereas the
remaining are not affected by any learning disorders. The subjects are university students or students that
finished university but that are still completing their academic formation with a master, a doctorate or
corporate training courses.
The test consisted in providing specific textbooks of 3 disciplines (political science, communication, and
economics) to students currently enrolled in affine degree, PhD or master courses, and ask them to study a
portion of a book each. 50\% of the students, equally divided in dyslexics and non-dyslexics, tried the RS that
was found to perform best, and obtained personalized suggestions about tools and strategies that should
help them in studying. The other 50\% received random recommendations, instead of the specific ones
output by the RS. Then, all the students were asked to follow the suggested methodologies during the study
phase and the level of knowledge of the studied disciplines was assessed by university professors, on a
scale from 0 to 10, 6 being the minimum sufficient score. A comparison was made between the results of the
ones that benefited of the RS and the ones that do not. Differences among dyslexic and non-dyslexic
students were also examined.

\section{Results}
\label{results}
In this section, the outcomes acquired from the experimentation will be presented. Initially, a comparison of the results obtained with the different weights used in the hybrid RS will be conducted. During this initial experiment, the optimal number of neighbors to be considered to compute the similarities will be also calculated. Additionally, this evaluation will incorporate the results obtained with the three different metrics employed for calculating similarities. Subsequently, we will perform a comparative analysis of the three considered approaches: user-based, item-based, and hybrid RSs.

Table \ref{tab:results} shows all the results obtained by the different configuration of the system, where n is the number of considered neighbors to compute the similarity and MAE\_x is the MAE obtained using the hybrid version with an $\alpha$ value equal to x. 

\begin{table}[h]
\centering
\begin{adjustbox}{width=\textwidth}
\begin{tabular}{|c|c|c|c|c|c|c|c|c|c|c|c|}
\hline
\textbf{similarity} & \textbf{n} & \textbf{MAE\_0} & \textbf{MAE\_1/8} & \textbf{MAE\_1/7} & \textbf{MAE\_1/6} & \textbf{MAE\_1/5} & \textbf{MAE\_1/4} & \textbf{MAE\_1/3} & \textbf{MAE\_1/2} & \textbf{MAE\_2/3} & \textbf{MAE\_1} \\ \hline
Euclidean           & 3                     & 1.1832          & 1.1315            & 1.1245            & 1.1152            & 1.1027            & 1.0846            & 1.0563            & 1.0049            & 0.9594            & 0.8920          \\ \hline
Euclidean           & 5                     & 1.1866          & 1.1347            & 1.1275            & 1.1179            & 1.1050            & 1.0863            & 1.0568            & 1.0015            & 0.9522            & 0.8750          \\ \hline
Euclidean           & 7                     & 1.1944          & 1.1412            & 1.1342            & 1.1250            & 1.1123            & 1.0939            & 1.0643            & 1.0085            & 0.9562            & 0.8735          \\ \hline
Euclidean           & 11                    & 1.2193          & 1.1650            & 1.1576            & 1.1479            & 1.1343            & 1.1146            & 1.0829            & 1.0244            & 0.9690            & 0.8782          \\ \hline
Cosine              & 3                     & 1.1839          & 1.1425            & 1.1371            & 1.1301            & 1.1212            & 1.1086            & 1.0895            & 1.0580            & 1.0309            & 0.9876          \\ \hline
Cosine              & 5                     & 1.2505          & 1.2020            & 1.1954            & 1.1867            & 1.1749            & 1.1576            & 1.1301            & 1.0815            & 1.0381            & 0.9647          \\ \hline
Cosine              & 7                     & 1.2754          & 1.2242            & 1.2177            & 1.2091            & 1.1972            & 1.1796            & 1.1511            & 1.0972            & 1.0486            & 0.9636          \\ \hline
Cosine              & 11                    & 1.3106          & 1.2550            & 1.2476            & 1.2380            & 1.2248            & 1.2054            & 1.1737            & 1.1136            & 1.0594            & 0.9659          \\ \hline
Pearson             & 3                     & 0.8217          & 0.8128            & 0.8118            & 0.8107            & 0.8096            & \cellcolor{yellow} \textbf{\large 0.8093}           & 0.8128            & 0.8314            & 0.8648            & 0.9661          \\ \hline
Pearson             & 5                     & 0.9092          & 0.8870            & 0.8843            & 0.8809            & 0.8766            & 0.8708            & 0.8639            & 0.8614            & 0.8735            & 0.9329          \\ \hline
Pearson             & 7                     & 0.9810          & 0.9494            & 0.9452            & 0.9398            & 0.9326            & 0.9227            & 0.9080            & 0.8913            & 0.8889            & 0.9222          \\ \hline
Pearson             & 11                    & 1.0868          & 1.0398            & 1.0336            & 1.0256            & 1.0147            & 0.9992            & 0.9756            & 0.9383            & 0.9160            & 0.9148          \\ \hline
\end{tabular}
\caption{Experimentation results considering the different similarity measures, number of neighbors and weights for the hybrid implementation. The best result (highlighted in the table) has been obtained by the hybrid method with an alpha value equals to 1/4, using Pearson as similarity measure and 3 neighbors.}
\label{tab:results}
\end{adjustbox}
\end{table}

Upon examining Table \ref{tab:results}, it can be seen that the hybrid model, utilizing an $\alpha$ value of 1/4, and employing Pearson's correlation with 3 neighbors, yielded the most favorable outcome with respect to the other analyzed cases. The achieved MAE by this model is 0.8093. To have a comparable measure, the relative error was calculated as

\begin{equation}    
    \varepsilon_r=\frac{1}{N}\sum_{i,j} (\frac{\left | p_{i,j} - q_{i,j} \right |}{q_{i,j}})
\end{equation}

where i indicates the i-th tool or strategy and j the j-th student tested, N is the total number of evaluations, $p_{i,j}$ is the real score given by the j-th student to the i-th methodology and $q_{i,j}$ is the real score given by the j-th student to the i-th methodology. The obtained result is 11.93\%.
 
There is currently no system that provides recommendations for this type of tools and study methodologies or anything similar. Nevertheless, the obtained error indicates the feasibility of utilizing the hybrid RS in suggesting the best supporting methodologies to dyslexic students.Indeed, an error of less than one point on the 5-point scale considered in the ratings was obtained. This implies that, for instance, when the RS proposes a methodology with a rating of 1 (indicating very limited usefulness), the actual rating would fall within the range of 0.2 (not useful at all) to 1.8 (somewhat useful), which would still be considered acceptable recommendations. Likewise, if the RS recommends a methodology with a rating of 4 (indicating high usefulness), this would imply that the actual rating would range between 3.2 (useful) and 4.8 (very much useful), again providing a reasonably accurate recommendation.

In addition, according to the reflected results, it appears that $n=3$ is the best number of neighbors for the algorithm, with an average MAE equals to 1.1816, whereas higher numbers of neighbors lead to worse performance in most cases.This behavior can be explained, since using a smaller number of neighbors allows for the recommendation of less popular items within the general population, thereby avoiding the spread of missinformative items~\citep{fernandez2021analysing}.

An additional noteworthy investigation within the scope of this research involved the determination of the optimal configurations for every single weight of the filter. To achieve this objective, each of the considered weight variants was subjected individual analysis, where the optimal configuration and MAE associated with each weight were identified. These results are summarized in Table \ref{tab:weights_analysis}.

\begin{table}[h]
\centering
\begin{tabular}{|c|c|c|}
        \hline
        \textbf{alpha} & \textbf{Best configuration} & \textbf{Best MAE}          \\ \hline
        0              & Pearson; n=3                & 0.8217                       \\ \hline
        1/8            & Pearson; n=3                & 0.8128                       \\ \hline
        1/7            & Pearson; n=3                & 0.8118                       \\ \hline
        1/6            & Pearson; n=3                & 0.8107                       \\ \hline
        1/5            & Pearson; n=3                & 0.8096                       \\ \hline
        \textbf{1/4}   & \textbf{Pearson; n=3 }      & \textbf{0.8093}              \\ \hline
        1/3            & Pearson; n=3                & 0.8128                       \\ \hline
        1/2            & Pearson; n=3                & 0.8313                       \\ \hline
        2/3            & Pearson; n=3                & 0.8648                       \\ \hline
        1              & Euclidean; n=7              & 0.8734                       \\ \hline
    \end{tabular}
\caption{Best configurations and MAEs obtained for the different weights in the hybrid model.}
\label{tab:weights_analysis}
\end{table}

\begin{figure}[ht]
    \centering
    \includegraphics[width=0.9\linewidth]{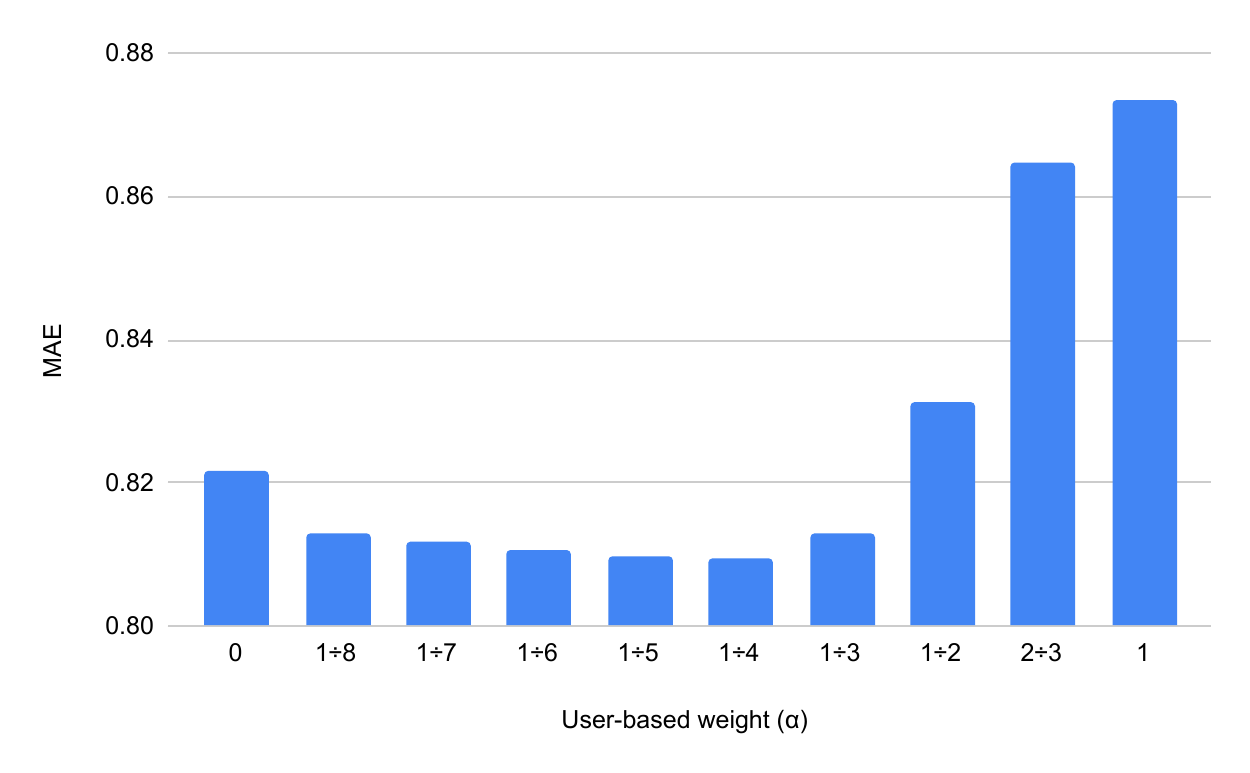}
    \caption{Best MAE for the different hybrid RS weight configurations.}
    \label{fig:bestConfChart}
\end{figure}

Another interesting observation from Table \ref{tab:weights_analysis} is that most of the time the best performance is achieved using the Pearson metric. Indeed, for each weight, the Pearson configuration consistently outperforms the others, except for the purely user-based filter, which is optimized using the Euclidean distance. Furthermore, in this case, it should be noted that the optimum number of neighbors is 7, instead of 3 as in the other cases. Nevertheless, the variations observed among the different results are not particularly substantial, spanning from 0.8734 to 0.8093. This disparity further diminishes when considering the hybrid models that prioritize the user-based methodology, where the MAE falls within the range of 0.8093 to 0.8217. Furthermore, it can be observed that an alpha value of 1/4 serves as the inflection point at which the system's performance begins to deteriorate in both directions, with a more pronounced decline as its value increases. These differences among the results of the optimal configurations for each weight are illustrated more clearly in Figure \ref{fig:bestConfChart}.

\begin{figure}[ht]
    \centering
    \includegraphics[width=\textwidth]{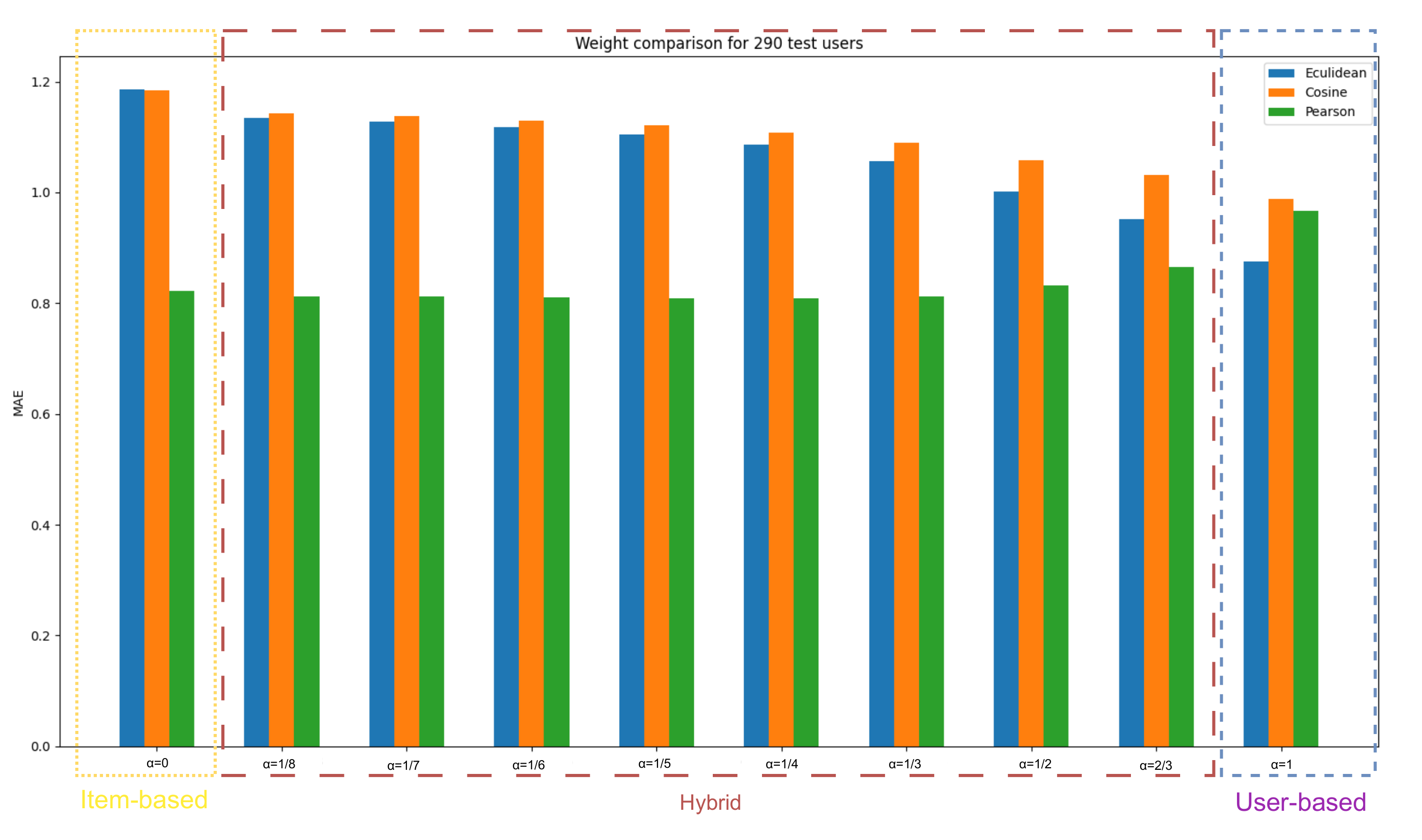}
    \caption{ Comparison among different weight-pairs of the hybrid approach according MAE.}
    \label{fig:Weight_comparison}
\end{figure}

The distinctions in performance among the different metrics are more evident in Figure \ref{fig:Weight_comparison}.
It shows the results obtained by the different similarity measures for the selected weights, considering the best performance of each metric. Among the similarities used, the Cosine distance had the highest average MAE (1.1432) considering all the different conducted experiments, indicating a poorer performance. The Euclidean distance yielded results similar to those of the Cosine distance, with an average MAE of 1.0723. Consequently, it may not be a suitable choice for our system. However, it's worth noting that, as demonstrated earlier, it performs better when applied in a user-based filter context. Finally, in comparison Pearson correlation exhibited lower MAE values being the best performance metric. It is also interesting to note that as the value of the parameter $\alpha$ increases, both Euclidean and Cosine similarities exhibit a decrease, while the Pearson similarity remains relatively stable until the point of transitioning to a fully user-based algorithm, at which juncture it undergoes an increase.

Since it achieved the lowest MAE, the hybrid model with $\alpha = 1/4$ using Pearson’s correlation and 3 neighbors to compute similarities was chosen as optimal to made recommendations about learning strategies and support tools to dyslexics student. To prove the efficacy of the procedure, this optimal recommendation system was also evaluated using precision and recall at k metrics~\citep{precision_recall}. In this context, a threshold of 3 on the rating scale was employed to deem a tool or strategy as relevant. This threshold was established by a panel of experts in methodologies for supporting dyslexic students, and simultaneously represents a typical threshold in literature when utilizing Likert-scale ratings within the 0 to 5 range \citep{JOORABLOO2021115485}. These metrics can be defined as follows:

\begin{itemize}
    \item Precision@k: is the proportion of recommended items in the top-k set that are relevant. It can be calculated as:
        \begin{equation}
            Precision@k = \frac{V@k}{R@k}
        \end{equation}
        where \textit{V@k} is the number of recommended items at k that are relevant for an user and \textit{R@k} is the total number of recommended items at k.
    \item Recall@k: is the proportion of relevant items found in the top-k recommendations. It can be obtained through: 
        \begin{equation}
            Recall@k = \frac{V@k}{T}
        \end{equation}
        where \textit{V@k} is the number of recommended items at k that are relevant for an user and \textit{T} is the total number of relevant items.
\end{itemize}

Similarly to the MAE calculation, precision@k and recall@k have been computed for the 290 users comprising the test set. Consequently, the average precision of the proposed hybrid recommendation system is 0.8524, while the average recall is 0.8278. According to precision, this signifies that 85\% of the methodologies recommended by the recommendation system to a specific student are relevant support tools or strategies for them. Furthermore, regarding the recall these results entail that the recommended methodologies by the system represent approximately 83\% of all tools relevant to that student. These results align with those obtained through MAE, reaffirming the utility of the proposed recommendation system.

Once the utility of the proposed hybrid recommendation model was verified, its optimal version was applied to assess the effectiveness of the algorithm in enhancing the learning experience of students in a real case. The experiment described in Section \ref{section4} was carried out and gave
the results reported in Table \ref{table5}, which show the average score obtained by the 4 groups (i) non-dyslexic
students that received suggestion by the RS, (ii) non-dyslexic students that do not received suggestion by
the RS, (iii) dyslexic students that received suggestion by the RS, (iv) dyslexic students that do not received
suggestion by the RS.

\begin{table}[h]
\caption{Score received by dyslexic and non-dyslexic students that adopted or not the RS system suggestions}
\begin{tabular}{l|c|c|}
\cline{2-3}
 & \multicolumn{1}{l|}{Received RS suggestions} & \multicolumn{1}{l|}{Not received RS suggestions} \\ \hline
\multicolumn{1}{|l|}{Non-dyslexic} & 8.6 & 8.2 \\ \hline
\multicolumn{1}{|l|}{Dyslexic} & 8.2 & 7.1 \\ \hline
\end{tabular}
\label{table5}
\end{table}

Such results clearly demonstrated how the adoption of the supporting tools and strategies suggested by the
implemented RS increases the score obtained by dyslexic students considerably (more than 1 point),
confirming that the proposed procedure can actually help them in their learning experience. In addition, also
non-dyslexic students showed an improvement of their average score (0.4 points), opening to interesting
scenarios for future research about the general usefulness of support methodologies suggested by AIs. A very outstanding result in our opinion is represented by the fact that, thanks to the RS, dyslexic students can
achieve the same performance of non-dyslexic students. This represents a step toward the real concept of
inclusivity.

\section{Conclusion}
\label{conclusion}
 Dyslexia can negatively impair multiple cognitive domains and has a high comorbidity with other disorders, therefore providing a customized support can be highly beneficial to people affected by it. In this work we wanted to explore the possibility of using machine learning, via recommendation systems, to suggest to dyslexic students the best study support methodologies in terms of learning tools and studying strategies in a personalized manner.

In order to perform this task, a collaborative-filtering recommendation system was chosen, evaluating the user-based, item-based and hybrid types, which have been adopted by different works in the education field. In addition three widely used metrics, namely the Euclidean distance, Cosine distance and Pearson distance, have been used in conjunction with the algorithms, to further increase their accuracies. The algorithms were trained, tested and validated on a self-evaluating questionnaire in which 1237 dyslexic students had to rate the most useful studying items among 17 support tools and 22 learning strategies. 

The results obtained show that the best performing recommendation system was able to recommend the most useful studying items to dyslexic students with an error of 11.93\%. This system exploits a hybrid filter with weight, $\alpha=1/4$ and a Pearson correlation as similarity metric.

To effectively evaluate the benefits of introducing a recommendation system to support students during their learning path, 50 students with and without dyslexia used the suggestions from the machine learning model whereas other students studied with randomized suggestions. The results show that the students using the recommendation system improved their performance, in particular dyslexic students obtained the highest benefits compared to non-dyslexic students.

These results obtained show that the recommendation system can be effectively used in education to support students with learning disorders during their education path.  However, further experiments are needed to further optimize the recommendation models for dyslexic students. For instance, a larger dataset will improve the analysis since it can increase the statistical power of our analysis. Moreover, recommendation systems can be applied to students without specific learning disorders in order to highlight possible differences in the learning tools used students without learning disorders and students with learning disorders. 
Indeed, these methodologies can significantly ameliorate the difficulties of students with specific learning disorders but can also benefit any student by providing a personalized way to approach learning.

\section{Acknowledgements}
Gianluca Morciano is a PhD student enrolled in the National PhD in Artificial Intelligence, XXXVII cycle, course on Health and life sciences, organized by Università Campus Bio-Medico di Roma.

José Manuel Alcalde Llergo is a PhD student enrolled in the National PhD in Artificial Intelligence, XXXVIII cycle, course on Health and life sciences, organized by Università Campus Bio-Medico di Roma.

These results are framed in VRAIlexia project funded by the European Union Committee within the Erasmus+ Programme 2014-2020 – Key Action 2: Strategic Partnership Project (Agreement n. 2020-1-IT02-K203-080006 – P.I.: Prof. Giuseppe Calabrò). 

\bibliography{rec_system_bib}

\end{document}